\documentclass[11pt]{article}
\usepackage{amssymb,cite}
\usepackage{graphicx}
\usepackage{latexsym}
\def\thefootnote{\fnsymbol{footnote}}
\newcommand{\eq}{\begin{equation}}
\newcommand{\en}{\end{equation}}
\newcommand{\be}{\begin{equation}}
\newcommand{\ee}{\end{equation}}
\newcommand{\eqa}{\begin{eqnarray}}
\newcommand{\ena}{\end{eqnarray}}
\newcommand{\ba}{\begin{eqnarray}}
\newcommand{\ea}{\end{eqnarray}}

\newcommand{\ZZ}{\hbox{{\rm Z{\hbox to 3pt{\hss\rm Z}}}}}

\newcommand{\Z}{\mathbb{Z}}
\newcommand{\Zp}{Z_{\mbox{\tiny{p}}}}
\newcommand{\Za}{Z_{\mbox{\tiny{a}}}}
\newcommand{\betag}{\beta_{\mbox{\tiny{g}}}}
\newcommand{\betas}{\beta_{\mbox{\tiny{s}}}}

\begin{document}
\begin{titlepage}
\vskip0.5cm
\begin{flushright}
DFTT 01/06\\
IFUP-TH 2006-01\\
DIAS-STP-06-01\\
\end{flushright}
\vskip0.5cm
\begin{center}
{\Large\bf  High precision Monte Carlo simulations of interfaces in the three-dimensional Ising model:}\\ 
{\Large\bf  a comparison with the Nambu-Goto effective string  model}
\end{center}
\vskip1.3cm
\centerline{
Michele~Caselle$^{a}$, Martin~Hasenbusch$^{b}$
 and Marco~Panero$^{c}$}
 \vskip1.0cm
 \centerline{\sl  $^a$ Dipartimento di Fisica
 Teorica dell'Universit\`a di Torino and I.N.F.N.,}
 \centerline{\sl Via Pietro~Giuria 1, I-10125 Torino, Italy}
 \centerline{\sl
e--mail: \hskip 1cm
 caselle@to.infn.it}
 \vskip0.4 cm
 \centerline{\sl  $^b$  
     Dipartimento di Fisica
     dell'Universit\`a di Pisa and I.N.F.N.,} 
   \centerline{ \sl Largo Bruno~Pontecorvo 3, I-56127 Pisa, Italy}

 \centerline{\sl
e--mail: \hskip 1cm
 Martin.Hasenbusch@df.unipi.it}
\vskip0.4 cm
 \centerline{\sl  $^c$ School of Theoretical Physics,
Dublin Institute for Advanced Studies,}
 \centerline{\sl
                  10 Burlington Road, Dublin 4,
                              Ireland}
 \centerline{\sl
e--mail: \hskip 1cm
 panero@stp.dias.ie}
 \vskip1.0cm
\begin{abstract}
Motivated by the recent progress in the effective string 
description of the interquark potential  in lattice 
gauge theory, we study interfaces with periodic boundary conditions
in the 
three-dimensional Ising model. Our Monte Carlo
results for the 
associated free energy are compared with the 
next-to-leading order (NLO) approximation of the Nambu-Goto string model. 
We find clear evidence for the validity of the effective string model at the level of the NLO truncation. 
\end{abstract}
\end{titlepage}

\setcounter{footnote}{0}
\def\thefootnote{\arabic{footnote}}
\section{Introduction}
\label{introsect}

Recently, the effective string description of the interquark potential in lattice gauge theories (LGT) has 
attracted a renewed interest~\cite{Caselle:2005vq,Dass:2005we,Billo:2005ej,Panero:2005iu,Caselle:2005xy,Meyer:2004hv,Majumdar:2004qx,Caselle:2004er,Caselle:2004jq,Juge:2004xr,Marescathesis,Koma:2003gi,Juge:2003vw,Majumdar:2002mr,Caselle:2002ah,Juge:2002br,Luscher:2002qv,Caselle:2002rm}: the increased computational power and improved algorithm efficiency~\cite{Luscher:2001up,Pepe:2001cx,deForcrand:2000fi} have allowed to perform stringent numerical tests of the model predictions, while a better understanding of the theoretical aspects was achieved~\cite{Billo:2005iv,Drummond:2004yp,Luscher:2004ib,Polchinski:1991ax,Billo:2006}. One of the plenary talks at the XXIII International Symposium on Lattice Field Theory held 
in Trinity College, Dublin, in July 2005 was devoted to the topic~\cite{Kuti:2005xg}.

Many of the mentioned studies are focused on the behaviour of 
the two-point Polyakov loop correlation function: 
the numerical results for the free energy associated with a pair of static external sources in a pure gauge 
theory 
were compared with predictions from effective string models, as a function of the interquark distance $r$ 
and of the inverse temperature $L$. 

One of the simplest effective string theories is based on the action originally formulated by Nambu and 
Goto~\cite{Goto:1971ce,Nambu:1974zg}: it is a purely bosonic model, which, despite the difficulties related 
to anomaly and non-renormalisability, has a straightforward geometric interpretation, and for which 
the leading order (LO) and next-to-leading order (NLO) terms 
in an expansion around the classical, long-string configuration 
agree with 
the effective model proposed by Polchinski and Strominger~\cite{Polchinski:1991ax}. Furthermore, 
the Nambu-Goto action also appears (together with other terms) in the derivation of an effective description 
for QCD~\cite{Koma:2002rw,Antonovthesis}. 

In this paper, as a further step in this direction, 
we compare the Nambu-Goto model with a set of high precision
results on the interface free energy of the Ising spin model in 
three dimensions with periodic boundary conditions in the directions
parallel to the interface. 
As it will be discussed in detail in section~\ref{isingsect}, an interface can 
be forced into the Ising spin system by introducing a seam of antiferromagnetic 
bonds through a whole cross-section of the system.  
Via duality, Wilson loops and Polyakov loop correlators in the $\Z_2$ gauge
theory are related with interfaces in the Ising spin model:
Wilson loops are mapped into interfaces with fixed boundary conditions in 
both directions, while  Polyakov loop correlators are mapped into interfaces 
with periodic boundary conditions in one direction and fixed boundary
conditions in the other direction. 
Hence, the present study is complementary to our previous work on the 
Polyakov-loop correlator; the periodic boundary conditions in both directions allow us 
to disentangle the pure string contributions from other effects, possibly (directly or indirectly) induced by fixed boundaries. 

Besides these reasons of interest, which are 
primarily motivated by the study of confinement in lattice gauge theories, fluid interfaces are also very interesting  
because they have
several experimental realizations, ranging from binary mixtures to amphiphilic membranes (for a review see for instance~\cite{GFP,Privman:1992zv}). Moreover the Nambu-Goto model 
(which is based on the assumption that the action of a given string configuration is proportional to the area of the surface spanned by the string during its time-like evolution) is closely related~\cite{Privman:1992zv} to the so-called 
capillary wave model~\cite{BLS}, which was proposed several decades ago (actually {\em before} the Nambu-Goto action), as a tool to describe interfaces in three-dimensional statistical physics systems. 

Finally, it is interesting to remark that interfaces in spin models are also naturally connected to  
maximal 't~Hooft loops in lattice gauge theory --- see, for instance,~\cite{deForcrand:2005pb,Bursa:2005yv}.

The problem of the interface with periodic boundary conditions has been studied in a number of articles, particularly 
in the early Nineties of last century --- see~\cite{Caselle:1994df,HaPi97} and references therein. 
The level of precision 
in these studies favoured the NLO prediction of the Nambu-Goto model against 
the Gaussian approximation, however it did not allow for a precise quantitative check of the NLO prediction itself.

The increase in computer power 
and a slightly smaller correlation length compared with~\cite{Caselle:1994df,HaPi97} allow
us a significantly better statistical control of the next to leading order corrections considered here. In particular, 
our present statistics is about 1000~times larger than that 
of~\cite{Caselle:1994df}.

The structure of this paper is the following: in section~\ref{isingsect} we briefly describe the introduction of an interface in the 3D Ising model, and the associated free energy; next, in section~\ref{algorithmsect}, we describe the numerical algorithm used in this work. Section~\ref{literaturesect} offers a review of known theoretical and numerical results, while our new results are presented in section~\ref{ourresultssect}. Finally, we summarise our conclusions in section~\ref{conclusionsect}.

\section{The 3D Ising model and the interface free energy}
\label{isingsect}

The confined phase of the $\Z_2$ gauge model in 3D is mapped by duality into the low temperature phase of the Ising spin model, 
where the global symmetry is spontaneously broken and
a  
non-vanishing magnetisation exists. According to the duality transformation,  
the observables of the gauge theory 
can be represented 
introducing an antiferromagnetic coupling for suitable sets of bonds in the spin model: these bonds pierce a surface having the original source lines as its boundary.    
This procedure can be naturally extended by introducing a seam of antiferromagnetic bonds throughout a whole time-slice on the lattice. Effectively, this amounts to imposing antiperiodic boundary conditions along the ``time-like'' direction, and 
induces an interface separating a  
domain of positive from a 
domain of negative magnetisation.\footnote{An 
alternative way to generate an interface in the Ising spin system would require to fix all the spin variables on the two opposite time-slices at the boundaries to the values $+1$ and $-1$, respectively; however such Dirichlet boundary conditions would lead to rather large finite-size effects. 
Further methods to determine the interface free energy are discussed in 
the literature. For a collection of articles on this subject see e.g.~\cite{wholevolume}.
}

These interfaces are the main subject of our analysis;  
to define the notations, we consider a periodic, cubic lattice with sizes 
$L_0 \times L_1 \times L_2$. Let $(x_0, x_1, x_2)$ denote the coordinates of the lattice sites, with $x_\mu = 0 , \dots , L_\mu$ for $ \mu = 0, 1, 2$. The action is given by:
\eq
\label{spinaction}
S = - \beta \sum_x \sum_{\mu}  J_{x,\mu} \; s_x s_{x+\hat \mu} \;\; . 
\en
Let us focus on two possible choices for the $J_{x,\mu}$ variables:  
\begin{itemize}
\item
setting $ J_{x,\mu}=1$ for all $x,\mu$, we obtain a system with periodic 
boundary conditions in all directions; 
the corresponding partition function is denoted by $\Zp$; 
\item
setting 
$J_{x,\mu}=-1$ for $x=(L_0,x_1,x_2)$
and $\mu=0$, and $J_{x,\mu}=+1$ for all the remaining links we obtain 
a system with antiperiodic boundary conditions in the $x_0$-direction and periodic
boundary conditions in the remaining directions; 
the corresponding partition function is denoted by $\Za$.
\end{itemize}
The latter choice induces an interface in the system, whose free energy\footnote{For convenience, we use the so-called ``reduced free energy'', which is dimensionless.} is given --- in first-approximation --- by the 
difference between the free energy of the system with anti-periodic and periodic 
boundary conditions: $F_s^{(0)} = -\ln \frac{\Za}{\Zp}$. 
This quantity has a 
characteristic $L_0$-dependence, 
due the fact that 
the interface induced by the anti-periodic boundary conditions 
enjoys full translational invariance in the $x_0$-direction. 
This entropy effect can be easily taken into account 
defining:
\begin{equation}
 F_s^{(1)} = -\ln \frac{Z_{\mbox{\tiny{a}}}}{Z_{\mbox{\tiny{p}}}} + \ln L_0 \;. 
\end{equation} 
However, for large values of $L_0$, an arbitrary odd (even) number of interfaces can appear in the system with antiperiodic (periodic) boundary conditions; 
assuming that the interfaces do not interact,\footnote{This assumption is reasonable for 
a low density of interfaces.} the reduced interface free energy can be defined as~\cite{Hasenbusch:1992eu}: 
\eq
\label{Fsdefinition}
F_s^{(2)} = \ln L_0 - \ln\left( \frac{1}{2} 
\ln\frac{1+Z_{\mbox{\tiny{a}}}/Z_{\mbox{\tiny{p}}}}{1-Z_{\mbox{\tiny{a}}}/Z_{\mbox{\tiny{p}}}}\right) \; ,
\label{fs2}
\en
which is used in the following. 

Finally, we would like to remark that the ratio of partition functions $Z_a/Z_p$ is 
directly related with the tunneling correlation length in a system with 
cylindrical geometry; for a detailed discussion see section 4.2
of \cite{Caselle:1994df}.

\section{The simulation algorithm}
\label{algorithmsect}

There are different methods available to compute the ratio of 
partition functions $Z_{\mbox{\tiny{a}}}/Z_{\mbox{\tiny{p}}}$ by Monte Carlo simulations. 

\begin{itemize}
\item
Integration of the energy difference $E_{\mbox{\tiny{a}}} - E_{\mbox{\tiny{p}}}$ over
$\betas$, starting from a value of $\betas$ in the high temperature phase of the 
Ising spin model, where the interface tension vanishes. 
(See e.g.~\cite{HaPi97} and references therein.)

\item
Snake-algorithm~\cite{deForcrand:2000fi,Pepe:2001cx}:  A sequence of systems that interpolate 
between the periodic and anti-periodic case is defined, introducing the defects one-at-a-time; the $Z_{\mbox{\tiny{a}}}/Z_{\mbox{\tiny{p}}}$ ratio can be factored as:
\eq
\label{factoredratio}
\frac{Z_{\mbox{\tiny{a}}}}{Z_{\mbox{\tiny{p}}}} = \frac{Z_{L_1 \times L_2}}{Z_{L_1 \times L_2 -1}} \cdot \frac{Z_{L_1 \times L_2-1}}{Z_{L_1 \times L_2 -2}} \cdot \dots \cdot \frac{Z_{1}}{Z_{0}} \; ,
\en
where $Z_k$ is the partition function associated with the system where $k$ defects have been introduced (so that $Z_{L_1 \times L_2}=Z_{\mbox{\tiny{a}}}$, while $Z_{0}=Z_{\mbox{\tiny{p}}}$). The free energy differences between $Z_k$ and $Z_{k+1}$ can be easily computed, as the two systems only differ by the value of $J_{x,\mu}$ on a single bond: since in general there is a sufficiently large overlap between configurations contributing to $Z_k$ and $Z_{k+1}$, importance sampling with respect to the denominators on the right-hand side of eq.~(\ref{factoredratio}) is possible. 
However, for any $0<k<L_1 \times L_2$ the translational invariance in the $x_0$-direction is broken. 
However, for $k$ sufficiently close to $L_1 \times L_2$, the entropy gain of ``depinning''
becomes comparable  with the energy cost and the interface starts to move along the lattice in the $x_0$-direction. This causes severe autocorrelation problems in a numerical simulation of these systems.

\end{itemize}
These two choices are quite similar in spirit: the free energy difference is evaluated from the sum of many small contributions that can be easily computed. Both methods 
allow to investigate large interfaces, and the computational effort required for a given precision grows only with a power of the lattice size. However, 
the obvious practical difficulty with both methods is that a large number of 
individual simulations have to be run. 

In the present work, we have measured the ratio of the partition functions 
$Z_{\mbox{\tiny{a}}}/Z_{\mbox{\tiny{p}}}$ directly, using 
a variant of the boundary flip algorithm~\cite{Hasenbusch:1992eu}. As in~\cite{Hasenbusch:1998gh}, we did not actually change the boundary conditions during the simulation: rather, we counted the configurations with periodic boundary conditions that would allow for this flip.  This method is efficient as long as $Z_{\mbox{\tiny{a}}}/Z_{\mbox{\tiny{p}}}$ is not too small. Since $Z_{\mbox{\tiny{a}}}/Z_{\mbox{\tiny{p}}} \simeq \exp(-\sigma L_1 L_2)$,  
$\sigma L_1 L_2 \lesssim 10$ is a rather strict upper bound on the interface size that can be reached with this method, since the signal to noise ratio decays
exponentially with the interface area.\footnote{Due to our enormous statistics we could obtain a meaningful result for interface areas with 
$\sigma L_1 L_2$ slightly larger than $10$.}
However, these lattice sizes are sufficient for our purpose as we shall see in the following. 

For the update of the configuration, we have used the standard single cluster algorithm~\cite{Wolff:1988uh}. For most of our simulations we have used the G05CAF random number generator of the NAG library, which is a linear congruential generator characterised by $a=13^{13}$, $c=0$ and modulus $m=2^{59}$. 
As a check of the reliability of the random number generator, we have 
repeated a few of the runs for $\betas=0.236025$ with the RANLUX generator
discussed in~\cite{Luscher:1993dy}, and 
the results are consistent with those 
obtained before with the G05CAF generator. Since there is no hint of a 
problem with the G05CAF, and the RANLUX generator is 
more time-consuming, we continued with the G05CAF generator.

\section{Summary of results given in literature}
\label{literaturesect}

In the following subsections we review the theoretical expectations and the numerical results which are available in the literature.

\subsection{Theoretical expectations}
\label{theoreticalsubsect}

A possible description for the dynamics of the interface in the continuum is provided by the Nambu-Goto model~\cite{Goto:1971ce,Nambu:1974zg}: it is based on the hypothesis that the action associated with a given interface configuration is proportional to the area of the interface itself:
\eq
\label{nambugoto}
S= \sigma \int d^2 \xi \sqrt{\det g_{\alpha\beta}} \;,
\en
where $\xi$ are the surface coordinates, while $g_{\alpha\beta}$ is the metric induced by the embedding in the 
three-dimensional space. For sake of simplicity, it is assumed that the interface can be parametrised in terms 
a single-valued, real function describing the transverse displacement of the surface with respect to a reference
 plane. This model is essentially the same as the capillary wave model~\cite{Privman:1992zv}, with the further 
 assumption that $\sigma$ does not depend on the direction of the normal to the infinitesimal surface element. 
 Here we neglect the theoretical difficulties associated with the fact that the 
model is actually anomalous, and non-renormalisable; in the following of the discussion, the model will be regarded as an effective theory expected to describe the dynamics of the interface for sufficiently large values of $\sigma L_1 L_2$ (\emph{i.e.} of the minimal interface area, in dimensionless units).

A perturbative expansion in powers of $(\sigma L_1 L_2)^{-1}$ yields the following result for the 
partition function associated with the interface: 
\be
\label{nloequation}
Z=\frac{\lambda}{\sqrt{u}}{\rm  e}^{-\sigma L_1L_2}
\Bigl| \eta\left(i u\right)/\eta\left(i\right) \Bigr|^{-2}
\left[1+\frac{f(u)}{\sigma L_1L_2}+O\left(\frac{1}{(\sigma L_1L_2)^2}
\right)\right]~~,
\label{2loop}
\ee
This expression was obtained for the first time in~\cite{df83} with a zeta-function regularization and then
re-obtained in~\cite{Caselle:1994df,cp96} with three other different types of regularization.
In eq.~(\ref{2loop}), $\lambda$ is a parameter that can be predicted by an argument from perturbation theory 
of the $\phi^4$ model in three dimensions (see below), $\tau=iu=iL_2/L_1$ is the modulus of 
the torus associated with the cross-section of the system, $\eta$ is Dedekind's function:
\be
\eta(\tau)=q^{1/24}\prod_{n=1}^{\infty}\left(1-q^n\right)~~,
\quad\quad q\equiv \exp(2\pi i \tau)~~,
\label{eta}
\ee
and
\be
f\left(u\right)=\frac{1}{2}\left\{
\left[\frac{\pi}{6} u E_2\left(i u\right)\right]^2 -
\frac{\pi}{6} u E_2\left(i u\right) + \frac{3}{4}\right\}
\label{fu}~~,
\ee
where $E_2(\tau)$ is the first Eisenstein series:
\begin{equation}
E_2(\tau)=1-24\sum_{n=1}^{\infty}\frac{n~ q^n}{1-q^n}~~,
\quad\quad q\equiv \exp(2\pi i \tau) \; ,
\label{e2}
\end{equation}
In most of our simulations, we have chosen $u=L_2/L_1=1$; for this choice
one gets: $f(1)=1/4$.

In the following we shall be interested in the interface free energy, which, for square lattices
of size $L_1=L_2\equiv L$ 
takes the form:
\eq
F_s=\sigma L^2 -\ln\lambda-\frac{1}{4\sigma L^2}+ O\left(\frac{1}{(\sigma L^2)^2}\right)
\en
This is the theoretical expectation which, in the following section,
 we shall compare with our numerical results for $F_s^{(2)}$ --- see eq.~(\ref{fs2}).

The value of $\lambda$ cannot be predicted by the effective interface model, however perturbation theory of the 3D $\phi^4$ model~\cite{Munster:1990yg,Provero:1995cz,HaPi97} gives: 
\begin{equation}
\ln \lambda = \frac{1}{2} \ln \sigma -\ln 2 + \ln S \; ,
\end{equation}
with:
\begin{equation}
S = \frac{4}{\sqrt{1 - \frac{u_r}{4 \pi} \left(\frac{39}{32} - \frac{15}{16} \ln 3 \right) }} \cdot  \frac{\Gamma(3/4)}{\Gamma(1/4)} \; .
\end{equation}
Using $u_r = 14.3(1)$~\cite{Caselle:1997hs}, 
one gets $G\equiv \ln 2 - \ln S \approx 0.29$.

\subsection{Numerical results}
\label{numericalsubsect}

In table~\ref{basictab} we have summarised numerical estimates
for basic quantities at the values of $\betas$ 
studied in the present work.  The result for the  
critical 
finite temperature
phase transition $N_t$ is taken from table~4 of~\cite{Caselle:1995wn}. 
The interface tension $\sigma$ is taken from table~8 of~\cite{Caselle:2004jq}. 
Note that in~\cite{Caselle:2004jq} only the leading order quantum corrections
were used to obtain these results. The system sizes were large enough to 
safely ignore NLO contributions. 
The result for the exponential correlation 
length $\xi$ is taken from table~1 of~\cite{Caselle:2004jq}. These numbers were obtained interpolating the results of~\cite{Agostini:1996xy,Caselle:1997hs} 
and from the analysis of the low temperature series~\cite{Arisue:1994wc}.
The magnetisation $m$ has been computed from the interpolation formula eq.~(10) of~\cite{Talapov:1996yh}.

\begin{table}
\caption{\sl \label{basictab} Summary of numerical estimates for basic quantities at the values of \emph{$\beta_{\mbox{\tiny{s}}}$} 
studied in this work. \emph{$\beta_{\mbox{\tiny{g}}}$} is the coupling 
of the $\Z_2$ gauge theory that corresponds, via duality, 
to the \emph{$\beta_{\mbox{\tiny{s}}}$} of the Ising spin model. $N_t$ is the inverse of the 
finite temperature phase transition, $\sigma$ is the interface tension, $\xi$
the bulk correlation length and $m$ the magnetisation. 
}
\begin{center}
\begin{tabular}{|l|l|c|l|l|l|}
\hline
\multicolumn{1}{|c}{$\betas$} &
\multicolumn{1}{|c}{$\betag$} &
\multicolumn{1}{|c}{$N_t$} & 
\multicolumn{1}{|c}{$\sigma$}  &
\multicolumn{1}{|c}{$\xi$} &
\multicolumn{1}{|c|}{$m$}\\
\hline
 0.276040  &  0.65608 & 2 & 0.204864(9)  & 0.644(1) &  0.85701  \\
 0.236025  &  0.73107 & 4 & 0.044023(3)  & 1.456(3) &  0.63407  \\
 0.226102  &  0.75180 & 8 & 0.0105241(15)& 3.09(1)  &  0.45311  \\
\hline
\end{tabular}
\end{center}
\end{table}

\section{New numerical results}
\label{ourresultssect}

In this section we present our new numerical results.

First we studied the finite $L_0$ effects in the interface free energy 
$F_s$ as defined by eq.~(\ref{Fsdefinition}). To this end, we run 
a series of simulations at $\betas=0.236025$ with $L_1=L_2$. 
Writing $Z_{\mbox{\tiny{a}}}$ and $Z_{\mbox{\tiny{p}}}$ in terms of eigenvalues of the 
transfer matrix 
(for discussion see e.g. section 4.2
of~\cite{Caselle:1994df}), one sees that the leading 
corrections in $L_0$ to $F_s(L_1,L_2)$ vanish as $\exp(-L_0/\xi)$.
Indeed, the results in table~\ref{R236025tab} show that
the results for $F_s(L_1,L_2)$ quickly
converge with increasing $L_0$. For all values of $L_1=L_2$ given in table~\ref{R236025tab}, the choice $L_0 = 3 L_1$ should guarantee that corrections 
due to the finiteness of $L_0$ are far smaller than the statistical errors of 
our numerical estimates. In the following, we shall use $L_0 = 3 L_1$ also for other values of $\betas$. 
In table~\ref{R027604tab} and in table~\ref{R226102tab} we present our results for 
$\betas=0.27604$ and $\betas=0.226102$, respectively. 

\begin{table}
\caption{\sl \label{R236025tab} Results for the interface free energy $F_s$ as
defined in eq.~(\ref{Fsdefinition}) at \emph{$\beta_{\mbox{\tiny{s}}}=0.236025$}. $L_0$, $L_1$ and $L_2$ give the linear sizes of the lattice and stat is the number of measurements. For each measurement, 10 single cluster updates were performed.
}
\begin{center}
\begin{tabular}{|c|c|c|c|l|}
\hline
  $L_0$ & $L_1$ &  $L_2$ &stat/100000  &\multicolumn{1}{|c|}{$F_s$}  \\
\hline
  6 & 6 & 6 &\phantom{0}500  & \phantom{0}3.37985(29) \\
  8 & 6 & 6 &\phantom{0}500  & \phantom{0}3.38689(26) \\
 10 & 6 & 6 &\phantom{0}500  & \phantom{0}3.38989(22) \\
 12 & 6 & 6 &\phantom{0}500  & \phantom{0}3.39067(20) \\
 18 & 6 & 6 &\phantom{0}500  & \phantom{0}3.39079(17) \\
\hline
  7 & 7 & 7 & \phantom{0}500&  \phantom{0}3.97243(37) \\  
 10 & 7 & 7 &           1000&  \phantom{0}3.99356(22) \\  
 14 & 7 & 7 & \phantom{0}500&  \phantom{0}3.99783(26) \\
 21 & 7 & 7 & \phantom{0}599&  \phantom{0}3.99802(19) \\  
\hline
 24 & 8 & 8 & \phantom{0}500 & \phantom{0}4.67900(28) \\
\hline
 18 & 9 & 9 &           1000 & \phantom{0}5.44197(35) \\ 
 27 & 9 & 9 &           1000 & \phantom{0}5.44170(28) \\ 
\hline
30 & 10 & 10 & \phantom{0}910& \phantom{0}6.29015(44) \\
\hline
 22 &11 &11 &          1000  & \phantom{0}7.22382(78) \\     
 33 &11 &11 &          1000  & \phantom{0}7.22219(64) \\  
\hline
 36 &12&12 &            1000 & \phantom{0}8.2441(10) \\
\hline
 26 &13 &13            &1000 & \phantom{0}9.3487(21) \\
 39 &13 &13            &1000 & \phantom{0}9.3481(17) \\
\hline
42 & 14 & 14&           1000 &           10.5403(30) \\
\hline
\end{tabular}
\end{center}
\end{table}

\begin{table}
\caption{\sl \label{R027604tab} Results for the interface free energy $F_s$ as
defined in eq.~(\ref{Fsdefinition}) at \emph{$\betas=0.27604$}. For all the sets of parameters $10^8$ measurements were performed; for each measurement 5 single cluster updates were performed.
}
\begin{center}
\begin{tabular}{|c|c|c|l|} 
\hline
  $L_0$ & $L_1$ &  $L_2$ & \multicolumn{1}{|c|}{$F_s$}  \\
\hline
 12 & 4 & 4 & \phantom{0}4.29672(23) \\
 15 & 5 & 5 & \phantom{0}6.18752(56) \\
 18 & 6 & 6 & \phantom{0}8.4669(16) \\
 21 & 7 & 7 &11.1540(57)  \\
 24 & 8 & 8 &14.239(25) \\
\hline
\end{tabular}
\end{center}
\end{table}

\begin{table}
\caption{\sl \label{R226102tab} Results for the interface free energy $F_s$ as
defined in eq.~(\ref{Fsdefinition}) at \emph{$\betas=0.226102$}. For each measurement 20 single cluster updates performed. In total, the simulations 
whose results are summarised in this table took about 2 years of CPU-time on a single PC with an Athlon XP 2000+ CPU. 
}
\begin{center}
\begin{tabular}{|c|c|c|c|l|}
\hline
  $L_0$ & $L_1$ &  $L_2$ &stat/100000 & \multicolumn{1}{|c|}{$F_s$}  \\
\hline
  30 & 10 & 10 &         1000 & \phantom{0}3.53042(11) \\  
  33 & 11 & 11 &         1000 & \phantom{0}3.78620(11) \\ 
  36 & 12 & 12 &         1000 & \phantom{0}4.05312(12) \\ 
  39 & 13 & 13 &         1000 & \phantom{0}4.33451(13) \\    
  42 & 14 & 14 &         1000 & \phantom{0}4.63149(15) \\   
  45 & 15 & 15 &         1000 & \phantom{0}4.94717(17) \\   
  48 & 16 & 16 &         1000 & \phantom{0}5.28138(19) \\ 
  51 & 17 & 17 &         1000 & \phantom{0}5.63492(22) \\ 
  54 & 18 & 18 &         1000 & \phantom{0}6.00959(27) \\   
  57 & 19 & 19 &         1000 & \phantom{0}6.40446(32) \\  
  60 & 20 & 20 &         1000 & \phantom{0}6.82040(38) \\ 
  63 & 21 & 21 &         1000 & \phantom{0}7.25587(46) \\ 
  66 & 22 & 22 &         1326 & \phantom{0}7.71339(50) \\
  69 & 23 & 23 &\phantom{0}999& \phantom{0}8.19094(72) \\    
  72 & 24 & 24 &         1033 & \phantom{0}8.68895(88) \\    
  75 & 25 & 25 &         1000 & \phantom{0}9.2063(12) \\
  78 & 26 & 26 &         1050 & \phantom{0}9.7462(14) \\
  81 & 27 & 27 &         1017 &           10.3062(19) \\
  84 & 28 & 28 &         1015 &           10.8894(25) \\ 
  87 & 29 & 29 &         1022 &           11.4881(33) \\
  90 & 30 & 30 &         1008 &           12.1181(45)\\
\hline
\end{tabular}
\end{center}
\end{table}

\subsection{Fitting the data}
\label{fittingdatasubsect}

\begin{figure}
\includegraphics[height=8cm]{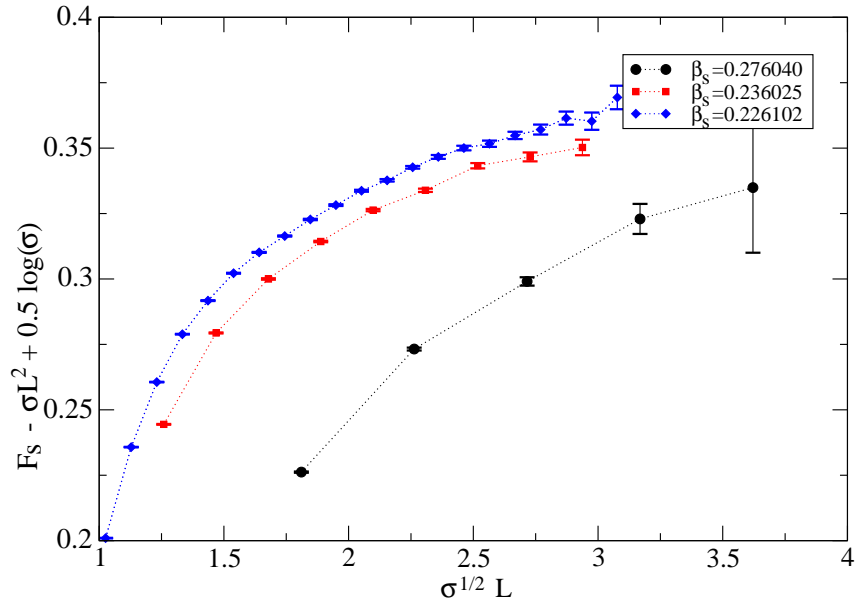}
\vskip0.5cm
\caption{$F_s - \sigma L^2 + 0.5 \ln \sigma$  as function of 
$\sqrt{\sigma} L$  for \emph{$\betas=0.226102$}, $0.236025$ and 
$0.27604$. In all cases $L_1=L_2=L$ and $L_0 = 3 L$. 
The values for $\sigma$ are taken from table~\ref{basictab}. We have 
checked that the plot does not change significantly, when $\sigma$ is 
varied within the quoted errorbars. The dotted lines should only guide the eye.
\label{figuresig}
}
\end{figure}

In figure~\ref{figuresig} we have plotted $F_s-\sigma L^2 + \ln(\sigma)/2$
as a function of the dimensionless quantity $\sqrt{\sigma} L$, 
where $L=L_1=L_2$.
The values for $\sigma$ are taken from 
table~\ref{basictab}. As $\betas \rightarrow \beta_c$, in the scaling limit,
the curves for different values of $\betas$ should fall on top of each other. While the curve for 
$\betas=0.27604$ is clearly different from the other two, those for $\betas=0.236025$ and $0.226102$ are close to each other --- their difference being approximately constant. 
We have checked that these observations still hold when varying $\sigma$ within the quoted errors. One should note that the LO effective string prediction corresponds
to $F_s-\sigma L^2 + \ln(\sigma)/2$ being constant.

Next we performed a more quantitative analysis of our data.
Motivated by the theoretical prediction of eq.~(\ref{2loop}), we fitted our data 
with the ansatz:
\begin{equation}
\label{fitsquare}
   F_s = \sigma L^2  + c_0 + \frac{c_2}{\sigma L^2}
\end{equation}
for the interface free energy. Using the interface tension 
$\sigma$ as parameter of the 
fit, we get results that are consistent with those in table~\ref{basictab}. However, the statistical error of our new results for $\sigma$
is clearly 
larger than the error quoted in table~\ref{basictab}.  Also, since we 
are mainly interested in the value of $c_2$, we have fixed $\sigma$ to the 
values given in table~\ref{basictab} in the following. 

Our results for the remaining fit parameters $c_0$ and $c_2$ are shown
in tables~\ref{fit2},~\ref{fit4} and~\ref{fit8} for 
$\betas=0.27604$, $0.236025$ and $0.226102$, respectively.  In these fits, 
we have included data for all available lattice sizes $L_1=L_2=L\ge L_{\mbox{\tiny{min}}}$.
Throughout, we have only included data obtained with $L_0=3 L$. 

\begin{table}
\caption{\sl \label{fit2}
Results of fits with the ansatz in eq.~(\ref{fitsquare}) of the data at
\emph{$\betas=0.27604$}. The interface tension has been fixed to $\sigma=0.204864$.
}
\begin{center}
\begin{tabular}{|c|l|l|c|}
\hline
$L_{\mbox{\tiny{min}}}$  & \multicolumn{1}{|c}{$c_0$}&\multicolumn{1}{|c}{$c_2$}& 
\multicolumn{1}{|c|}{$\chi^2/$d.o.f.} \\
\hline
 4 & 1.1500(14) &  -0.430(5) & 0.81 \\
 5 & 1.154(5)   &  -0.45(3)  & 0.92 \\
\hline
\end{tabular}
\end{center}
\end{table}

\begin{table}
\caption{\sl \label{fit4} 
Results of fits with the ansatz in eq.~(\ref{fitsquare}) of the data at
\emph{$\betas=0.236025$}. The interface tension has been fixed to $\sigma=0.044023$.
}
\begin{center}
\begin{tabular}{|c|l|l|c|}
\hline
$L_{\mbox{\tiny{min}}}$  & \multicolumn{1}{|c}{$c_0$}&\multicolumn{1}{|c}{$c_2$}& 
\multicolumn{1}{|c|}{$\chi^2/$d.o.f.} \\
\hline
  7   & 1.9320(5) &  -0.1972(13) &  4.22 \\
  8   & 1.9348(8) &  -0.208(3)   &  2.37 \\
  9   & 1.9383(13)&  -0.223(5)   &  0.86 \\
 10   & 1.9387(22)&  -0.225(11)  &  1.00 \\
 11   & 1.944(4)  &  -0.258(22)  &  0.65 \\
 12   & 1.935(7)  &  -0.195(48)  &  0.27 \\
\hline
\end{tabular}
\end{center}
\end{table}

\begin{table}
\caption{\sl \label{fit8} 
Results of fits with the ansatz in eq.~(\ref{fitsquare}) of the data at
\emph{$\betas=0.226102$}. The interface tension has been fixed to $\sigma=0.0105241$.
}
\begin{center}
\begin{tabular}{|c|l|l|c|}
\hline
$L_{\mbox{\tiny{min}}}$  & \multicolumn{1}{|c}{$c_0$}&\multicolumn{1}{|c}{$c_2$}& 
\multicolumn{1}{|c|}{$\chi^2/$d.o.f.} \\
\hline
 16   &   2.65513(51)  &  -0.1855(16)  &     7.62 \\
 17   &   2.65853(65)  &  -0.1994(24)  &     2.49 \\
 18   &   2.66067(85)  &  -0.2090(34)  &     1.33 \\
 19   &   2.6627(11)   &  -0.2187(49)  &     0.73 \\
 20   &   2.6636(15)   &  -0.2240(72)  &     0.70 \\
 21   &   2.6656(20)   &  -0.235(10)   &     0.52 \\
 22   &   2.6642(27)   &  -0.227(15)   &     0.50 \\
\hline
\end{tabular}
\end{center}
\end{table}

To estimate the effect of the error in $\sigma$, we redid the fits
for  $\betas=0.236025$ with $\sigma=0.04402$. This leads to slightly smaller
values of $c_2$, e.g. for $L_{\mbox{\tiny{min}}}= 10$ we get $c_2=-0.227(11)$. 
We also repeated the fit for $\betas=0.226102$ using
$\sigma=0.0105226$ as input for the interface 
tension. This leads to a slight decrease in $c_2$; for instance, for $L_{\mbox{\tiny{min}}}=22$ 
we get $c_2 = -0.232(15)$ instead of $c_2 = -0.227(15)$ for 
$\sigma=0.0105241$; the error on the value of $\sigma$ that is used 
as input in the fits only plays a minor role.

The result $c_2 \approx - 0.45$ at $\betas=0.27604$ is clearly inconsistent with the 
prediction $c_2=-0.25$. However, we should note that $\xi < 1$ and we should expect huge scaling corrections. 

The fit results for $\betas=0.236025$ and $\betas=0.226102$ have similar features.
In both cases, the value of $c_2$ increases as $L_{\mbox{\tiny{min}}}$ is increased. Also
the $\chi^2/$d.o.f. decreases as $L_{\mbox{\tiny{min}}}$ is increased.
For $\betas=0.236025$,  $\chi^2/$d.o.f. $\approx 1$ is reached at $L_{\mbox{\tiny{min}}}=9$,
where $c_2=-0.223(5)$. For the slightly larger $L_{\mbox{\tiny{min}}}=11$ we get:
$c_2=-0.258(22)$, which is fully consistent with the theoretical prediction.

For $\betas=0.226102$, $\chi^2/$d.o.f. $\approx 1$ is reached at $L_{\mbox{\tiny{min}}}=19$, where $c_2=-0.2187(49)$. For $L_{\mbox{\tiny{min}}}=21$ we get: $c_2=-0.235(10)$, which is consistent with the theoretical prediction within two units of the standard deviation. 

Next we fitted our data for $\betas=0.226102$  with the ansatz:
\begin{equation}
\label{fitsquare2}
F_s= \sigma L^2 + c_0 + \frac{c_2}{\sigma L^2} + \frac{c_4}{(\sigma L^2)^2}
\end{equation}
to check possible effects of higher order corrections on the numerical results
for $c_0$ and $c_2$.  The results are displayed in table~\ref{fit8bis}. Again, we have checked that the error of the input value for $\sigma$ is not relevant. Now the numerical results for $c_2$ are smaller than the theoretical prediction $c_2=-0.25$. Adding higher order corrections to the fit does not allow for a more accurate numerical determination 
of $c_2$. However these fits clearly show that the small deviation of $c_2$ obtained from the fit to eq.~(\ref{fitsquare}) can be explained by higher order corrections that are omitted in the ansatz. 

\begin{table}
\caption{\sl \label{fit8bis}
Fit results for the ansatz in eq.~(\ref{fitsquare2}) of the data for $F_s$ at
\emph{$\betas=0.226102$}. The interface tension has been fixed to $\sigma=0.0105241$.
}
\begin{center}
\begin{tabular}{|c|l|l|l|c|}
\hline
$L_{\mbox{\tiny{min}}}$  & \multicolumn{1}{|c}{$c_0$}&\multicolumn{1}{|c}{$c_2$}&
\multicolumn{1}{|c}{$c_4$}&
\multicolumn{1}{|c|}{$\chi^2/$d.o.f.} \\
\hline
      14    &   2.6629(10) & -0.244(6) & 0.103(9) & 4.48 \\
      15    &   2.6696(14) & -0.293(9) & 0.187(15)& 1.03 \\
      16    &   2.6731(20) & -0.320(15)& 0.240(26)& 0.55 \\
      17    &   2.6713(27) & -0.305(22)& 0.207(42)& 0.52 \\
      18    &   2.6716(37) & -0.308(32)& 0.214(70)& 0.57 \\
\hline
\end{tabular}
\end{center}
\end{table}

Finally, let us briefly discuss the results for $c_0$. The results are quite 
stable for different values of $L_{\mbox{\tiny{min}}}$. Also, fits to eq.~(\ref{fitsquare})
and eq.~(\ref{fitsquare2}) give consistent results. 
As a final result, we quote $c_0=1.154(5)$, $1.944(5)$ and $2.665(5)$ for 
$\betas=0.27604$, $0.236025$ and $0.226102$, respectively.
Correspondingly, one gets: $c_0+\frac{1}{2} \ln \sigma = 0.361(5)$, $0.382(5)$ and $0.388(5)$, 
which is somewhat larger than the theoretical prediction $G \approx 0.29$ from~\cite{Munster:1990yg,HaPi97}.

\subsection{Results for $L_1 \ne L_2$}
\label{nonsquaresubsect}

For $\betas=0.236025$ we have also performed simulations for lattices with non-square cross-section ($L_1 \ne L_2$): the results of these simulations are given in table~\ref{R236025tabasym}. In order to compare with the theoretical prediction 
for the NLO contribution to $F_s$, we have subtracted the classical and the leading order contribution from $F_s$. To this end, we have used $\sigma=0.044023$ from table~\ref{basictab} and  $c_0=1.944$ from the fits summarised in table~\ref{fit4}. For comparison, in the last column of table~\ref{R236025tabasym} we give the theoretical prediction for the NLO contribution --- see eq.~(\ref{nloequation}). The absolute value of the numerical results is found to be about $10 \%$ larger than the theoretical 
prediction for the NLO contribution. This can be interpreted as an effect due to higher order corrections.
It is interesting to observe that such higher order terms become more and more important as the ratio
$L_2/L_1$ increases: this is an effect we already observed in our previous analysis of the Polyakov loop
correlators~\cite{Caselle:2005vq}.

\begin{table}
\caption{\sl \label{R236025tabasym} Results for the interface free energy $F_s$ as
defined in eq.~(\ref{Fsdefinition}) at \emph{$\betas=0.236025$}. 
Results for $L_1 \ne L_2$. We also give $F_s-$ LO, where we have used 
$\sigma=0.044023$  and $c_1 = 1.944$ as input.  
}
\begin{center}
\begin{tabular}{|c|c|c|c|l|l|l|}
\hline
 $L_0$ & $L_1$ &  $L_2$ &stat/100000  &\multicolumn{1}{|c|}{$F_s$} &
\multicolumn{1}{|c|}{$F_s-$ LO}  & \multicolumn{1}{|c|}{NLO} \\
\hline
36 &  10 & 12 &           1000 & \phantom{0}7.1670(6) &--0.0489(6) &--0.0487\\
45 &  10 & 15 & \phantom{0}693 & \phantom{0}8.4449(12)&--0.0471(12)&--0.0440\\
54 &  10 & 18 &          1000  & \phantom{0}9.6976(17)&--0.0493(17)&--0.0439\\
60 &  10 & 20 &          1029  &           10.5235(25)&--0.0518(25)&--0.0454\\ 
66 &  10 & 22 &\phantom{0}999  &           11.3466(36)&--0.0521(36)&--0.0478\\  
\hline
\end{tabular}
\end{center}
\end{table}

\section{Conclusions}
\label{conclusionsect}

In this work, we studied interfaces in the 3D Ising model with periodic boundary conditions.  
We compared our numerical results for the interface free energy with 
predictions derived from the Nambu-Goto effective string model,
which is essentially equivalent to the capillary wave model. 
Compared with previous work on the problem~\cite{Caselle:1994df}, the statistical errors are reduced by a factor of about 30, which allows for a quantitative 
check of the next-to-leading order (\emph{i.e.} beyond the free string approximation)
prediction.
For the two coupling values closest to the phase transition, we found for a linear extension 
$\sqrt{\sigma} L \gtrapprox 2$ of the interface a good 
agreement with the next-to-leading order prediction of the Nambu-Goto
model. Expressed in terms of the inverse deconfinent temperature $N_t$
this corresponds to $L \gtrapprox 2.5 N_t$. 
In the case of the 
Polyakov loop correlation function we 
found in~\cite{Caselle:2005vq} a similar behaviour 
{\em along the compactified direction of the Polyakov loop correlator}. 
On the contrary, along the direction with Dirichlet boundary conditions 
clear deviations from the Nambu-Goto 
string prediction were observed for distances of the order of $2.5 N_t$.
In fact, we actually found  that the Nambu-Goto string fits the numerical data
for the interquark potential at low temperatures less well than its free
string approximation.
Even if the presence of a boundary term in the effective string 
action is 
ruled out
(at least in three dimensions)   by
string duality arguments~\cite{Luscher:2004ib}\footnote{Notice however that the argument of\cite{Luscher:2004ib}
only rules out the dominant boundary correction (the one which can be reinterpreted as a shift in the interquark
distance~\cite{Caselle:2004jq}) but it does not exclude possible higher order boundary corrections. Notice also that the
absence of such a dominant boundary corrections is compatible with the numerical data (see 
again~\cite{Caselle:2004jq}) on short distance Polyakov loop correlators.}, the above comparison between the
 present results and our previous analysis on the Polyakov loop correlators clearly shows that some kind 
of boundary correction is present in the latter case.  

By virtue of the absence of boundary effects, we think that the interface
free energy   
discussed in this paper is the perfect setting to study 
limits and merits of effective string models and also, if possible, to improve these effective descriptions.    
In this respect it would be very  interesting to further analyze the 
deviations from the Nambu-Goto predictions which we observe in
the range $\sqrt{\sigma}L<2$.
 Understanding the origin of these deviations
remains one of the most intriguing challenges   
towards
a consistent and satisfactory effective string
description of the confining potential in lattice gauge theories.

\vskip1.0cm {\bf Acknowledgements.} This work was partially supported by the European Commission TMR programme HPRN-CT-2002-00325 (EUCLID). M.~P. acknowledges support received from Enterprise Ireland under the Basic Research Programme. The authors would like to thank M.~Bill\'o, D.~Ebert, L.~Ferro and F.~Maresca for useful discussions.

\end{document}